\documentclass[aip,apl] {revtex4-1} 
\usepackage{graphicx}  
\usepackage{dcolumn}   
\usepackage{bm}        
\usepackage{amssymb}   
\usepackage{dcolumn}
\usepackage[novolumeabbr]{unitsdef} 
\usepackage{epsfig}
\newunit{\invcm}{\centi\meter\unitsuperscript{-1}}

\begin{document}

\title{Ultrafast carrier phonon dynamics in NaOH-reacted graphite oxide film}

\author{Dongwook Lee}
\affiliation{Division of Physics and Applied Physics, Nanyang Technological University, 637371, Singapore}
\affiliation{Cavendish Laboratory, University of Cambridge, Cambridge CB3 0HE, United Kingdom}
\email{dongwookleedl324@gmail.com}
\author{Xingquan Zou}
\affiliation{Division of Physics and Applied Physics, Nanyang Technological University, 637371, Singapore}
\author{Xi Zhu}
\affiliation{Division of Materials Science, School of Materials Science and Engineering, Nanyang Technological University, Singapore 639798}
\author{J. W. Seo}
\affiliation{Division of Physics and Applied Physics, Nanyang Technological University, 637371, Singapore}
\affiliation{Cavendish Laboratory, University of Cambridge, Cambridge CB3 0HE, United Kingdom}
\affiliation{School of Advanced Materials Science and Engineering, Sungkyunkwan University, Suwon 440-746, Republic of Korea}
\author{Jacqueline M. Cole}
\affiliation{Cavendish Laboratory, University of Cambridge, Cambridge CB3 0HE, United Kingdom}
\affiliation{Department of Chemistry, University of New Brunswick, P. O. Box 4400, Fredericton, NB, E3B5A3, Canada}
\affiliation{Department of Physics, University of New Brunswick, P. O. Box 4400, Fredericton, NB, E3B5A3, Canada}
\author{Federica  Bondino}
\affiliation{IOM-CNR, Laboratorio TASC, S.S.14, Km.163.5, I-34149 Basovizza, Italy}
\author{Elena Magnano}
\affiliation{IOM-CNR, Laboratorio TASC, S.S.14, Km.163.5, I-34149 Basovizza, Italy}
\author{Saritha K. Nair}
\affiliation{Division of Physics and Applied Physics, Nanyang Technological University, 637371, Singapore}
\author{Haibin Su}
\affiliation{Division of Materials Science, School of Materials Science and Engineering, Nanyang Technological University, Singapore 639798}

\date{\today}
\begin{abstract}
NaOH-reacted graphite oxide film was prepared by decomposing epoxy groups in graphite oxide into hydroxyl and -ONa groups with NaOH solution. Ultrafast carrier dynamics of the sample were studied by time-resolved transient differential reflection ($\triangle$R/R). The data show two exponential relaxation processes. The slow relaxation process ($\sim$2ps) is ascribed to low energy acoustic phonon mediated scattering. The electron-phonon coupling and first-principles calculation results demonstrate that -OH and -ONa groups in the sample are strongly coupled. Thus, we attribute the fast relaxation process ($\sim$0.17ps) to the coupling of hydroxyl and -ONa groups in the sample.
\end{abstract}

\maketitle

Graphite comprises stacked layers of graphene, which is a two-dimensional atomic structure of carbon atoms covalently bonded with $sp^{2}$-orbitals in a hexagonal lattice. Graphite loses its electrical conductivity after being oxidized into graphite oxide (GO). GO also has a layered structure but has \textit{two} kinds of carbon atoms: $sp^{2}$-hybridized and $sp^{3}$-hybridized carbon atoms\cite{1,2,3}. The coexistence of $sp^{2}$- and $sp^{3}$-hybridized carbon atoms leads to the warping and bending of the associated carbon plane. Various chemical groups such as epoxy (-O-) and hydroxyl (-OH) groups are also present on the surface of GO \cite{1,4,6}. Although the optical and electrical properties of graphene and graphite have been investigated with pump-probe techniques \cite{7,8}, the optical properties of GO-related materials have not been fully investigated. For this work, we slightly modified the structure of GO by preparing a NaOH-GO film, which is devoid of epoxy groups. We report the carrier phonon dynamics of the NaOH-GO film. Studying carrier dynamics of NaOH-GO film will lead to better understanding of the carriers' behavior in graphene, graphite, GO, and GO-related materials.

\begin{figure}[!b]
\epsfig{file=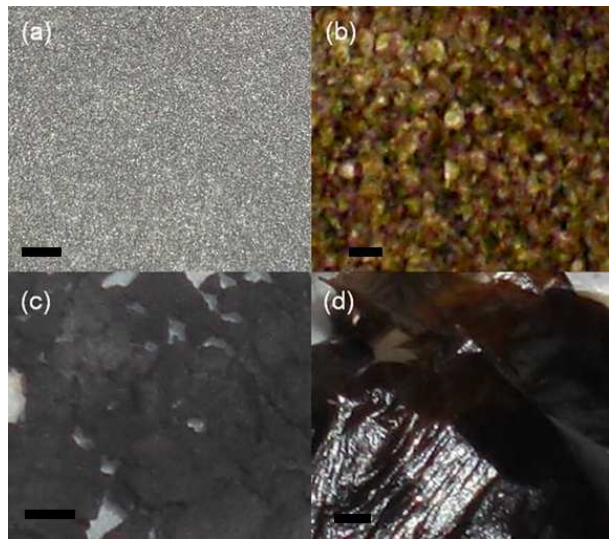, width=8.0cm}
\caption{(Color online) Sample pictures. (a) graphite, (b) GO, (c) GO reacted with NaOH, and (d) NaOH-GO film. The scale bar in (a) - (d) is 1mm.}
\end{figure}

GO was prepared by the Brodie method \cite{6,9}. NaOH-GO film was synthesized by immersing GO powders into 0.1M NaOH solution for 2 weeks, washing them with deionized water, and drying them in the air \cite{10}. The NaOH-GO film was characterized by Fourier transform infrared spectroscopy (FT-IR, Thermo Nicolet Avatar 360 spectrometer), X-ray Photoemission Spectroscopy (XPS), and Pump-probe optical reflectivity measurements. XPS experiments were conducted on the BACH beamline at Elettra in Italy. A degenerate pump-probe setup \cite{11} was employed with a Ti:sapphire mode-locked laser with 80 MHz pulse repetition rate, generating 60fs pulses at 800 nm. The pump beam was focused on the sample with a spot size of 60 $\mu$m to generate photoexcited carriers, while the weaker 30$\mu$m-diameter probe beam measured the change in reflectivity ($\triangle$R/R) as a function of the probe delay time relative to the pump. The intensity ratio of the pump pulse (0.0375 nJ/pulse) and probe pulse is $\sim$30, while the polarization between the pump (P polarized) and probe (S polarized) is orthogonal in order to minimize detection of the pump signal. The pump beam was chopped at a frequency of 1 MHz, and the $\triangle$R/R at different pump-probe delay times was measured with a lock-in amplifier (Stanford Research System SR844 RF).

\begin{figure}[!t]
\epsfig{file=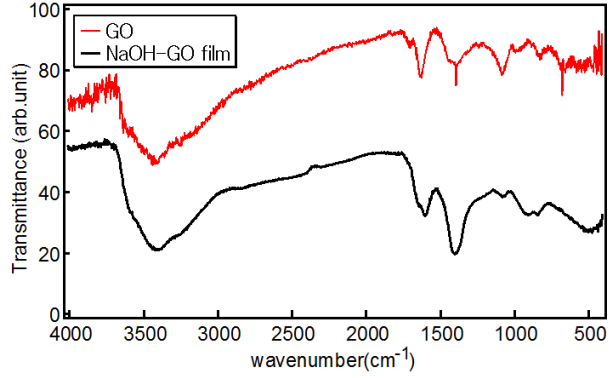, width=8.0cm}
\caption{(Color online) FT-IR spectra of GO and NaOH-GO film.}
\end{figure}

\begin{figure}[!b]
\epsfig{file=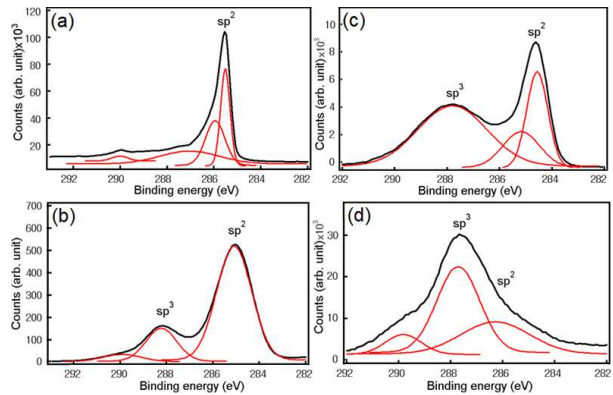, width=8.0cm}
\caption{(Color online) C core-level XPS spectra of (a) polycrystalline graphite (99.995+\%, Aldrich), (b) GO, (c) NaOH-reacted GO, and (d) NaOH-GO film.}
\end{figure}

Figure 1(a) shows graphite, which is black and conductive. After oxidation, it turns brown (Figure 1(b)). GO turns black after reaction with NaOH (Figure 1(c)). NaOH-GO film (Figure 1(d)) which is flexible looks dark brown. Unlike GO, which has two kinds of carbon atoms, most of the carbon atoms in the NaOH-GO film are bonded through $sp^{3}$-hybridization \cite{10}. Figure 2 illustrates FT-IR spectra of GO and NaOH-GO film, which have four main peaks centered at 1050, 1380, 1650, and 3470 \invcm, representing epoxy groups, C-O vibrational mode, ketone groups, and hydroxyl groups (O-H stretching mode), respectively. Most epoxy groups in GO decompose into -ONa and -OH groups during the reaction with NaOH; the peak intensity at 1380 \invcm increases as a result of break-up of epoxy groups, while the peak at 1050 \invcm from epoxy groups diminishes.

The C core-level XPS spectrum of graphite in Figure 3(a) shows that carbon atoms in graphite have only $sp^{2}$-hybridized orbitals \cite{12}. The peak centered at 285.4 \eV is assigned to carbon-carbon (C-C) bonds in aromatic networks in graphite and has a well-known asymmetric line shape. The peak at 285.8 \eV refers to a defect peak such as anthracene and bisanthene \cite{13}. The origin of a broad peak at 287.1\eV is not clear. A plasmon peak is also present near 289.8 \eV \cite{14,15}. In Figure 3(b), GO has three peaks: 285.3 \eV, 288.4 \eV, and 290.4 \eV. The first peak is from C-C which exhibits $sp^{2}$-hybridized bonding \cite{12}. The second peak originates from C-O in alcohol which is bonded via $sp^{3}$-hybridized orbitals \cite{17,18,19}. The third peak is from ($\pi \rightarrow \pi^{*}$) excitation \cite{16}. NaOH-reacted GO has three main peaks (Figure 3(c)), which are centered at 284.6 \eV, 285.2 \eV, and 287.8 \eV. NaOH can decompose epoxy groups into -ONa and -OH. Therefore, more hydroxyl groups were produced after GO reacted with NaOH. In comparison with GO (Figure 3(b)), NaOH-reacted GO shows a large peak from $sp^{3}$-hybridized orbitals at 287.8 \eV (Figure 3(c)); this confirms that the peak at 287.8 \eV must arise from C-O in alcohol. The peak at 285.2 \eV in Figure 3(c) can be assigned to structural defects with $sp^{2}$-hybridized orbitals. The reaction with NaOH makes GO more defective. Meanwhile, NaOH-GO film (Figure 3(d)) has three peaks centred at 286.3 \eV, 287.7 \eV, and 289.8 \eV. The peak at 286.3 \eV might be due to unreacted double bonding which manifests $sp^{2}$-hybridized orbitals. The peak at 287.7 \eV originating from C-O in alcohol becomes sharper. The peak at 289.8 \eV represents carbon atoms of -COOH. The XPS peaks in Figure 3 were fitted with Gaussian functions to determine their areas. The ratio of $sp^{2}$ to $sp^{3}$ is extracted by dividing these areas. The $sp^{2}$:$sp^{3}$ ratio for GO, NaOH-reacted GO, and GO-film is 4.0, 0.8, and 0.3, respectively.

\begin{figure}[!t]
\epsfig{file=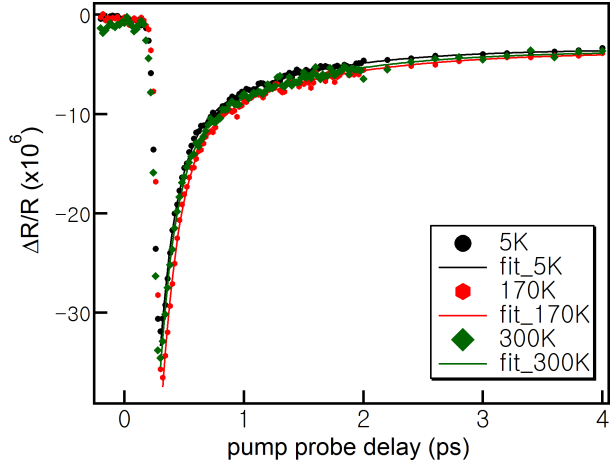, width=8.0cm}
\caption{(Color online) Differential reflectivity ($\triangle$R/R) versus pump-probe delay of NaOH-GO film at different temperatures.}
\end{figure}

The differential reflectivity ($\triangle$R/R) of the NaOH-GO film is shown in Figure 4 at several temperatures. The sign of $\triangle$R/R is negative and the recovery of $\triangle$R/R exhibits two relaxation processes. The data were fitted with a double-exponential decaying function, $\triangle$R/R = $A_{fast}$*exp(-t/$\tau_{fast}$) + $A_{slow}$*exp(-t/$\tau_{slow}$). The negative sign of $\triangle$R/R and the two relaxation processes are also found in bilayer graphene grown by chemical vapor deposition (CVD) \cite{6}. Ultrafast studies \cite{20,21,22,23, Dawlaty} show that carrier dynamics in graphene, few-layered graphene (FLG), and graphite are explained by two relaxation times ($\tau_{fast}$ and $\tau_{slow}$), and these two relaxation processes are ascribed to the optical phonon emission ($\tau_{fast}$) and acoustic-phonon mediated phonon scattering process  ($\tau_{slow}$). By fitting the data ($\triangle$R/R) of NaOH-GO film at various temperatures, we find that the two relaxation processes after pump excitation are in the similar time scales to those of pristine graphene, FLG and GO \cite{7, 24,25,26}. However, the $\tau_{fast}$ and $\tau_{slow}$ in NaOH-GO film are temperature-independent, as shown in Figure 5. Since $\tau_{slow}$ is related to the acoustic-phonon mediated scattering process after the pump excitation, its value is dependent on the lattice temperature. In NaOH-GO film, the lattice temperature is inhomogeneous due to the strong disorder in the structure, which also leads to the similar portion of local structure with similar high temperature regardless of the variation of lattice temperature. Thus, $\tau_{slow}$ exhibits the temperature-independent feature in the NaOH-GO film in comparison with the temperature dependent one in graphene and few-layered graphene. $\tau_{fast}$ of graphene, FLG, and reduced GO is of the order of ~100fs and is ascribed to carrier-optical phonon interaction\cite{8, 23, Dawlaty, Huang, 24}. However, $\tau_{fast}$ of NaOH-GO film is $\sim$180fs, which is almost double the value of graphene, FLG, and reduced GO. Moreover, it is confirmed in Figure 2 that the C-O vibration mode of epoxy groups has little contribution in the NaOH-GO film. Thus, $\tau_{fast}$ of the NaOH-GO film has a different relaxation route from graphite compounds including epoxy groups.


\begin{figure}[!t]
\epsfig{file=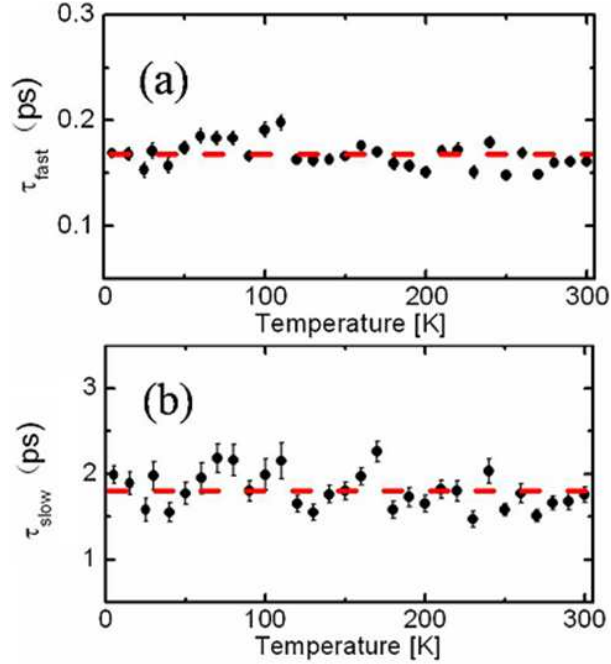, width=8.0cm}
\caption{(Color online) The two relaxation processes in NaOH-GO film as a function of lattice temperature. (a) temperature-independent $\tau_{fast}$. (b) temperature-independent $\tau_{slow}$. The red solid lines are linear fits, and the dashed lines are guides to the eyes.}
\end{figure}

The NaOH-GO film has many defects such as wrinkles, distortions, and surface groups. Among them, randomly distributed -OH and -ONa groups play an important role in the ultrafast carrier dynamics of the NaOH-GO film. Figure 6 schematically illustrates the structure of the NaOH-GO film, which has wrinkles, distortions, and many chemical groups such as -OH and -ONa groups on the layer surfaces. The -OH and -ONa groups in NaOH-GO film have two kinds of stretching modes: intraplanar and interplanar vibration. These act, respectively, like dangling bonds swinging freely on the layers or vibrating between the layers. In addition, hydrogen-bond formation between interlayers accelerates -OH group vibrations. The functional group vibration can be coupled to generate phonons in amorphous materials \cite{27,28,29}.

\begin{figure}[!t]
\epsfig{file=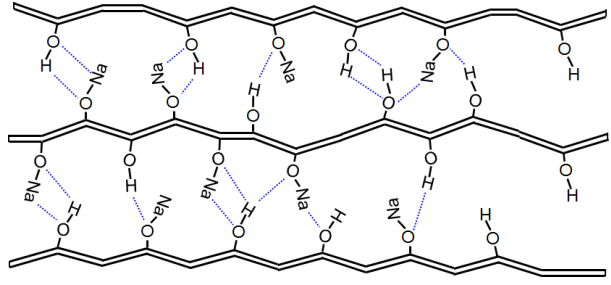, width=8.0cm}
\caption{(Color online) Schematic diagram of a NaOH-GO film. The layers are wrinkled and have so many chemical groups. Hydrogen bonds are apparent between chemical groups (dot lines).}
\end{figure}

To understand the fast relaxation processes, first-principles calculations were performed, based on four types of GO-film structures containing: (A) only epoxy groups (Figure 7(a)); (B) ketone groups exclusively (Figure 7(b)); (C) two surface layers -OH groups, each on opposite GO face (Figure 7(c)); (D) surface layers -OH and -ONa groups, each on opposite GO faces (Figure 7(d)). These four structure types were constructed based on our experimental FT-IR and XPS observations. Density functional theory was employed within the local density approximation (LDA) using Quantum ESPRESSO [30] code with separable norm-conserving pseudopotentials and a plane-wave basis set. A kinetic energy cut-off of 60 \textit{Ry} and 8$\times$8$\times$8 \textit{k}-points were employed. Each atomic structure is fully relaxed in phonon and electron phonon coupling (EPC) calculations (\textit{q} mesh), until forces acting on atoms are less than 0.0001\eV/{\AA}. The EPC matrix elements have been calculated in the first Brillouin zone (BZ) on s 4$\times$4$\times$4 \textit{q} mesh obtained by employing Gaussian smearing of 0.015 \textit{Ry} on a 8$\times$8$\times$8 \textit{k} mesh. .

\begin{figure}[!t]
\epsfig{file=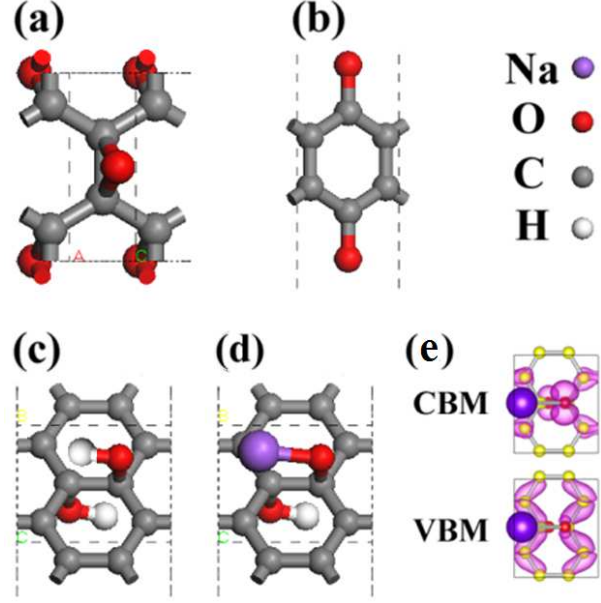, width=8.0cm}
\caption{(Color online) Atomic structures for four GO structures concerned with this work, containing (a) only epoxy groups, (b) only ketone groups, (c) -OH groups on the two surfaces, and (d) -OH and -ONa groups on the two surfaces. The purple, red, gray and white colors represent sodium, oxygen, carbon and hydrogen atoms, respectively. In addition, (e) shows $\Gamma$ point wavefuction plot for the electron doping of Figure 7(d). Yellow, purple, and red circles represent carbon, sodium, and oxygen atoms, respectively.}
\end{figure}

The electron-phonon coupling strength associated with a specific phonon mode and wavevector $\lambda_{nk}$ is given by \cite{31}:

Equation 1
\begin{equation}
\lambda_{nk} = \sum_{m,\nu} \int\frac{dq}{A_{BZ}}|g_{mn\nu}(k,q)^{2}|[\frac{n_{q\nu}+1-f_{mk+q}}{(E_{F}-\varepsilon_{mk+q}-\omega_{q\nu})^{2}}+\frac{n_{q\nu}+f_{mk+q}}{(E_{F}-\varepsilon_{mk+q}+\omega_{q\nu})^{2}}]
\end{equation}

Here $\varepsilon_{nk}$ is the energy eigenvalue of an electron with band index \textit{n} and wavevector \textit{k}, while $\omega_{q\nu}$ is the energy eigenvalue of a phonon with branch index $\nu$ and wavevector \textit{q}. $E_{F}$ is the Fermi energy. $A_{BZ}$ is the area of the first Brillouin zone where the integration is performed. $f_{nk}$ and $n_{q\nu}$ are the Fermi-Dirac and the Bose-Einstein factors, respectively. $g_{mn,\nu}$=$\langle mk+q|\Delta V_{q\nu}(r)|nk\rangle$ is the scattering amplitude of an electronic state $|nk\rangle$ into another state $|mk+q\rangle$ resulting from the perturbation in the self-consistent field potential $\Delta V_{q\nu}(r)$ arising from a phonon with branch index $\nu$ and wavevector \textit{q}. $\lambda$ at the $\Gamma$ point was focused in our calculation. We applied 1\% electron doping for the EPC calculations. The strongest EPC occurs at 27.04THz (37fs, $\lambda$=1.06) in A and at 44.7THz (22fs, $\lambda$=0.52) in B. The strongest EPC for the -OH and -OH in C occurs at 6.5THz (153fs, $\lambda$=0.0524). For structure D, the EPC is strongest at 5.49THz (182fs) with a lambda value 1.83. This is the same as the value of $\tau_{fast}$ obtained by the pump-probe measurements. Thus, we attribute the origin of the $\tau_{fast}$ to the coupling of -OH and -ONa groups. We also applied a 1\% hole doping for the EPC calculation. However, the strong EPC remains unchanged regardless of the amount of hole doping. Since Equation 1 includes the Bose-Einstein factor, the electron-phonon coupling strength for crystalline materials is temperature-dependent. In Figure 3, the ratio of $sp^{2}$ to $sp^{3}$- in NaOH-GO film is 0.3, indicating that the material has very strong disorder character. Therefore, the temperature-dependent carrier-phonon dynamics feature is suppressed substantially by the strong disorder in the NaOH-GO film and the values of $\tau_{fast}$ and $\tau_{slow}$ become temperature-independent. The wavefunctions at valance band maximum (VBM) and conduction band minimum (CBM) at the $\Gamma$ point of the structure D are plotted in Figure 7(e). The VBM wavefunction lies at the carbon with delocalized character, and CBM wavefunction is mainly localized near oxygen in O-H and O-Na bonds. Both wavefunctions will couple to the phonon modes consisting of the O-H and O-Na bonds. The relatively localized feature of CBM wave function results in more substantial coupling with the motion associated with hydroxyl and -ONa, which is dominant in the electron doping situation. Moreover, the relatively heavy mass of sodium accounts for the longer time scale of this fast process as compared with that of graphene, FLG, and reduced GO.

We modified the structure of GO by removing epoxy groups and prepared a NaOH-GO film. We characterized its structure with FT-IR and XPS. To investigate its carrier-phonon dynamics, we performed pump-probe measurements at different temperatures. $\tau_{fast}$ and $\tau_{slow}$ are temperature-independent. According to the first-principles calculation, the vibrational modes of -OH and -ONa groups in the samples are coupled. This coupling is so stable that they interact through hydrogen bonding between interlayers. We attribute $\tau_{fast}$ to the carrier-phonon coupling involved with -OH and -ONa groups in NaOH-GO film.

D. W. Lee acknowledges Cambridge Overseas Trust and Peterhouse College for financial support. J. M. Cole is indebted to the Royal Society for a University Research Fellowship and the University of New Brunswick, Canada, for the UNB Vice-Chancellor's Research Chair.

\clearpage


\newpage
\begin{thebibliography}{40}


\bibitem{1} D. W. Lee, L. de Los Santos, J. W. Seo, L. L. Felix, A. Bustamante D., J. M. Cole, and C. H. W. Barnes, \textit{J. Phys. Chem. B} \textbf{114}, 5723 (2010).

\bibitem{2} D. W. Lee and J. W. Seo, \textit{J. Phys. Chem. C} \textbf{115}, 2705 (2011).

\bibitem{3} W. Scholz and H. P. Boehm, \textit{Z. Anorg. Allg. Chem.} \textbf{369}, 327 (1969).

\bibitem{4} A. Lerf, H. He, M. Forster, and J. Klinowski, \textit{J. Phys. Chem. B} \textbf{102}, 4477 (1998).


\bibitem{6} D. W. Lee and J. W. Seo, \textit{J. Phys. Chem. C} \textbf{115}, 12483 (2011).

\bibitem{7} X. Zou, D. Zhan, X. Fan, D. W. Lee, S. K. Nair, L. Sun, Z. Ni, Z. Luo, L. Liu, T. Yu, Z. Shen, and E. E. M. Chia, \textit{Appl. Phys. Lett.} \textbf{97}, 141910 (2010).

\bibitem{8} X. Zhao, Z. Liu, W. Yan, Y. Wu, X. Zhang, Y. Chen, and J. Tian, \textit{Appl. Phys. Lett.} \textbf{98}, 121905 (2011).

\bibitem{9} B. C. Brodie, \textit{Philos. Trans. R. Soc.} \textbf{149}, 249 (1859).

\bibitem{10} D. W. Lee, J. W. Seo, G. R. Jelbert, L. de Los Santos V., J. M. Cole, C. Panagopoulos, and C. H. W. Barnes, \textit{Appl. Phys. Lett.} \textbf{95}, 172901 (2009).

\bibitem{11} E. E. M. Chia, D. Talbayev, J. X. Zhu, J. D. Thompson, A. J. Taylor, H. Q. Yuan,T. Park, C. Panagopoulos, G. F. Chen, J. L. Luo, and N. L. Wang, \textit{Phys. Rev. Lett.} \textbf{104}, 027003 (2010).

\bibitem{12} J. D\'{\i}az, G. Paolicelli, S. Ferrer and F. Comin, \textit{Phys. Rev. B} \textbf{54}, 8064 (1996).

\bibitem{13} H. Estrade-Szwarckopf, \textit{Carbon} \textbf{42}, 1713 (2004).

\bibitem{14} E. Desimoni, G. I. Casella, A. Morone and A. M. Salvi, \textit{Surf. Interface Anal.} \textbf{15}, 627 (1990).

\bibitem{15} F. L. Coffman, R. Cao, P. A. Pianetta, S. Kapoor, M. Kelly and L. J. Terminello, \textit{Appl. Phys. Lett.} \textbf{69}, 568 (1996).

\bibitem{16} E. Papirer, R. Lacroix, J. B. Donnet, G. Nanse and P. Fioux, \textit{Carbon} \textbf{32}, 1341 (1994).

\bibitem{17} T. T. Cheung, \textit{J. Appl. Phys.} \textbf{53}, 6857 (1982).

\bibitem{18} H. A. Katzman, P. M. Adams, T. D. Le and C. S. Hemminger, \textit{Carbon} \textbf{32}, 379 (1994).

\bibitem{19} H. Ago, T. Kugler, F. Cacialli, W. R. Salaneck, M. S. P. Shaffer, A. H. Windle and R. H. Friend, \textit{J. Phys. Chem. B} \textbf{103}, 8116 (1999).

\bibitem{20} R. W. Newson, J. Dean, B. Schmidt, and H. M. van Driel, \textit{Opt. Express} \textbf{17}, 2326 (2009).

\bibitem{21} M. Breusing, C. Ropers, and T. Elsaesser, \textit{Phys. Rev. Lett.} \textbf{102}, 086809 (2009).

\bibitem{22} T. Kampfrath, L. Perfetti, F. Schapper, C. Frischkorn, and M. Wolf, \textit{Phys. Rev. Lett.} \textbf{95}, 187403 (2005).

\bibitem{23} P. A. George, J. Strait, J. Dawlaty, S. Shivaraman, M. Chandrashekhar, F. Rana, and M. G.  Spencer, \textit{Nano Lett.} \textbf{8}, 4248 (2008).

\bibitem{24} B. A. Ruzicka, L. K. Werake, H. Zhao, S. Wang, and K. P. Loh, \textit{Appl. Phys. Lett.} \textbf{96}, 173106 (2010).

\bibitem{Dawlaty} J. M. Dawlaty, S. Shivaraman, M. Chandrashekhar, F. Rana, and M. G. Spencer, \textit{Appl. Phys. Lett.} \textbf{92}, 042116 (2008).

\bibitem{25} H. N. Wang, J. H. Strait, P. A. George, S. Shivaraman, V. B. Shields, M. Chandrashekhar, J.  Hwang, F. Rana, M. G. Spencer, C. S. Ruiz-Vargas, and J. Park, \textit{Appl. Phys. Lett.} \textbf{96}, 081917 (2010).

\bibitem{26} W. K. Tse and S. D. Sarma, \textit{Phys. Rev. B} \textbf{79}, 235406 (2009).

\bibitem{Huang} L. Huang, G. V. Hartland, L. Chu, Luxmi, R. M. Feenstra, C. Lian, K. Tahy, and H. Xing, \textit{Nano Lett.} \textbf{10}, 1308 (2010).



\bibitem{27} C. A. Murray and T. J. Greytak, \textit{Phys. Rev. B} \textbf{20}, 3368 (1979).

\bibitem{28} R. B. Laughlin and J. D. Joannopoulos, \textit{Phys. Rev. B} \textbf{17}, 4922 (1978).

\bibitem{29} E. Balan, S. Delattre, M. Guillaumet, and E. K. H. Salje, \textit{Am. Mineral.} \textbf{42}, 1257 (2010).

\bibitem{30} P. Giannozzi, S. Baroni, N. Bonini, M. Calandra, R. Car, C. Cavazzoni, D. Ceresoli, G. L. Chiarotti, M. Cococcioni, I. Dabo, et al., \textit{J. Phys.: Condens. Matter} \textbf{21}, 395502 (2009).

\bibitem{31} G. Grimvall, The Electron-Phonon Interaction in Metals; North Holland: Amsterdam 1981.

\end{thebibliography}
\end{document}